# Correlations for a new Bell's inequality experiment

Louis Sica

Code 5630
Naval Research Laboratory
Washington, D. C.  20375 USA
(202)-767-9466
FAX: (202)-767-9203
e-mail: sica@ccs.nrl.navy.mil

Proofs of Bell's theorem and the data analysis used to show its violation have commonly assumed a spatially stationary underlying process. However, it has been shown recently that the appropriate Bell's inequality holds identically for cross correlations of three or four lists of $\pm 1$'s, independently of statistical assumptions. When data consistent with its derivation are analyzed without imposition of the stationarity assumption, the resulting correlations satisfy the Bell inequality.

1.  **INTRODUCTION**

It has recently been shown [1,2] that Bell's inequalities depend on only one of the several assumptions historically employed in their derivation: namely, that cross correlations are performed among three or four lists of data, as appropriate to the inequality, each datum restricted to $\pm 1$. Bell's inequalities are identically satisfied by such data lists independently of the physical assumptions usually believed to underlie them. From closely related reasoning [2], it follows that Bell's inequality necessarily constrains the single function characterizing the correlation of periodic, spatially stationary stochastic processes, and that such processes cannot produce the cosine correlation of the singlet state.

Processes that are both more general than spatially stationary, and interesting from the perspective of Bell correlation characterization might initially be thought to constitute an empty set. However, the purpose of the present paper is to show that when the data of the customary real experiments are ordered so as to be consistent with the *derivation* of Bell's inequality, the set of correlation functions that result conforms to the definition of a spatially non-stationary process. This paper will be concerned mainly with relating real experimental data to Bell's inequality. However, in the final section, a strategy will be sketched for dealing with the correlations of intrinsically imaginary data, or counterfactuals, deriving from Bell's gedankenexperiment. It has been found that non-stationary correlations emerge here also, but their detailed discussion will be reserved for a separate paper.

A few words must be devoted to relating the developments presented here to prior work. Over time, Bell's inequalities have been re-derived from a more general perspective involving fewer assumptions. Peres [3], Redhead [4], and especially Eberhard [5], have given derivations based on ideas that overlap those presented in [1,2]. Eberhard in particular, demonstrated that Bell's inequality is a purely mathematical result. However, the implications of this fact as reported herein have not been described previously as far as the author is aware. Fine [6], derived Bell's inequalities from probability theory alone. However, from the perspective of experiments and their necessarily finite data, the derivation in [1,2] has greater generality than can be obtained from probability considerations, and has lead to new results. Recently, other authors [7,8] have articulated a view similar to that described in [1,2].

## 1.1 Bell's Theorem

To better discuss the issues and conceptual revisions implied in the new derivation of the Bell theorem [1,2], it is useful to review Bell's derivation [9-11] to identify its basic assumptions, not all of which have been sufficiently noted. The physical setting for the theorem is the Bohm version [12] of the EPR gedankenexperiment [13] schematized in Fig. 1. In this experiment, identical particles in a singlet state fly out from a source in different directions and encounter two Stern-Gerlach spin measuring apparatuses oriented at angles $\theta_A$ and $\theta_B$. These produce random readings $A(\theta_A)$ and $B(\theta_B)$. Particles emerge from this apparatus in two discrete directions, indicating a spin of $\pm 1$ in units of $\hbar/2$. When $\theta_A = \theta_B$, the measured values of $A$ and $B$ are always equal and opposite. This follows from properties of the singlet state, which has the same form for every (equal) angular setting of the spin meters.

Bell deduced from quantum mechanics that if the spin-meters pointed in arbitrary directions $\theta_A$ and $\theta_B$, not both the same, then the correlation of the output readings is given by

$$\langle A(\theta_A)B(\theta_B)\rangle = -Cos(\theta_A - \theta_B). \qquad (1)$$

He was not able to create a local model for this correlation which is defined as one based on the assumption that the spin meters act independently on particles prepared from common initial conditions at a past time when the singlet state was formed. However, he found that he could account for the correlation if information was transmitted instantaneously from one spin measuring device to the other at the time of measurement. (Such information is defined as nonlocal because it travels faster than the signal velocity of light, and is not attenuated with distance.) Bell sought to generalize this finding through a theorem that shows that no local model for the cosine correlation exists. That is why it could not be found.

Bell assumed deterministic solutions for the spin measurement problem of the form

$$A(a, \lambda) = \pm 1, \quad B(b, \lambda) = \pm 1, \qquad (2)$$

where $a = \theta_A$, $b = \theta_B$, and $\lambda$ designates a collection of random variables e.g., initial conditions, with associated probability density $\rho(\lambda)$. The locality assumption is represented in (2) by the fact that $A$ depends only on $a$, the information indicating its own angular setting, and $B$ depends only on $b$. That the correlation of spin values equals 1 at equal detector angles is ensured by the condition $A(a, \lambda) = -B(a, \lambda)$. Thus, one function determines all readouts and correlations of readouts, as $\lambda$ takes on its random values. A function $A(a, \lambda)$ with these properties is said to define a stochastic process [14].

In terms of function $A(a, \lambda)$, correlation $\langle AB \rangle$ is defined by

$$P(a,b) = -\int d\lambda \rho(\lambda) A(a, \lambda) A(b, \lambda). \qquad (3)$$

Intrinsic to the process so far specified is the fact that measurement values exist at multiple angular settings of $A$ and $B$ for each $\lambda$ i.e. for each realization of $A(a, \lambda)$. Thus, the value of any third readout exists (except at zero crossings). This readout $A(c,\lambda)$ is said to be "counterfactual" because it cannot be read from single particle pairs for which only one spin measurement per particle is possible. Bell ruled out additional spin meters acting after $A$ and $B$ [15]. Their sequential readings would not necessarily corresponding to the same value of  . And since the hidden variables  are unknown, assuming that they exist, the difficulty cannot be circumvented by repetition of the experiment to measure the spin at another instrument setting. Thus, the values assigned to the counterfactuals and their correlations depend on theoretical considerations and the model assumed for the underlying random process.

It must be observed that the assumption that $A(c,\lambda)$ may be read at *any number* of different values of angle $c$ for a given  represents a dramatic violation of quantum mechanical principles. This is due to the fact that the operators representing $A(a)$ and $A(a')$ for the same particle do not commute. Noncommutation in quantum mechanics is interpreted to mean that a sequence of operations to measure $A(a), A(a'), A(a)$ does not necessarily return the same value for $A(a)$ in both instances which is unlike the case of the stochastic process defined above. However, measurement $A(a)$ commutes with $B(b)$ since these represent operations on two different particles.

Thus, for the stochastic process Bell defined, he could compute a difference of correlations involving real and counterfactual variables:

$$P(a,b) - P(a,c) = - \int d\lambda \rho(\lambda) \times [A(a,\lambda)A(b,\lambda) - A(a,\lambda)A(c,\lambda)] . \quad (4)$$

By (2), $A^2 = 1$ for any setting, and by factoring,

$$P(a,b) - P(a,c) = \int d\lambda \rho(\lambda) A(a,\lambda) A(b,\lambda)[A(b,\lambda)A(c,\lambda) - 1] . \quad (5)$$

*This factoring step is pivotal in the construction of the Bell inequality. For each value of $\lambda$, the same value $A(a,\lambda)$ multiplies both $A(b,\lambda)$ and $A(c,\lambda)$. Three lists of data, of infinite length in general, are generated as $\lambda$ traverses its values, and the correlations of the inequality are the cross correlations of these three data lists.* After taking the absolute value of (5) one has

$$|P(a,b) - P(a,c)| \leq \int d\lambda \rho(\lambda)[1 - A(b,\lambda)A(c,\lambda)] , \quad (6)$$

or

$$|P(a,b) - P(a,c)| - P(b,c) \leq 1 . \quad (7)$$

Upon inserting quantum mechanical correlations

$$P(a,b) = -cos(\theta_a - \theta_b), \quad P(a,c) = -cos(\theta_a - \theta_c),$$
$$P(b,c) = -cos(\theta_b - \theta_c), \quad (8)$$

it is found that (7) is violated at certain angles. Note that the correlations of real data pairs, and the correlations of counterfactual and real data pairs, are assumed to be given by the same (quantum mechanical) correlation function depending only on the angular differences of detector settings. These additional assumptions specialize the random process still further. A stochastic process with second order correlations given by a single function of coordinate differences is said to be "stationary in the wide sense" [14]. It is noted that by adding an additional fourth variable, the same assumptions yield a similar derivation of the four variable inequality [16].

Several assumptions have now been identified in the construction of the Bell theorem. A question may be raised as to their necessity. It may be answered by re-deriving the inequality with fewer assumptions.

## 1.2 New Derivation of Bell's Inequality

Assume the existence of three lists, *a, b,* and *b'* of length $N$ composed of elements $a_i, b_i, b'_i$, each equal to $\pm 1$. From the $i$ th elements of the three lists form

$$a_i b_i - a_i b'_i = a_i (b_i - b'_i) = a_i b_i (1 - b'_i b_i) \quad . \tag{9}$$

Sum (9) from *1* to *N*, and divide by *N*. Then take the absolute value of both sides to obtain

$$\left| \sum_{i=1}^{N} a_i b_i / N - \sum_{i=1}^{N} a_i b'_i / N \right| \le \sum_{i=1}^{N} |a_i b_i| |1 - b'_i b_i| / N = \sum_{i=1}^{N} (1 - b'_i b_i) / N \tag{10}$$

From inspection of (9) and (10), it is seen that the algebraic form of the inequality depends on a factoring step just as was the case in (4) and (5). Finally,

$$\left| \sum_{i=1}^{N} a_i b_i / N - \sum_{i=1}^{N} a_i b'_i / N \right| \le 1 - \sum_{i=1}^{N} b'_i b_i / N \quad . \tag{11}$$

(A similar method may be used to derive an inequality from the assumption of four lists.) Thus, if three lists of data, items restricted to $\pm 1$'s, are produced by an experiment, the cross correlations of the lists satisfy Bell's inequality identically and independently of fluctuations about correlation means. The inequality follows as a mathematical result independent of physics, or restriction to a particular type of random process, or even the assumption of randomness. The result was discovered from an inquiry as to whether the Bell inequality holds under conditions of statistical fluctuation. *Clearly it holds independently of such fluctuation.* If the numerical correlation estimates in (11) approach ensemble average limits, as $N \to \infty$, then replacing the estimates in (11) with these limits results in the usual form of Bell's inequality.

The above derivation separates the purely mathematical facts underlying Bell's inequality from the explicit and implicit physical assumptions used in the usual derivation. The logical precision thereby acquired leads to new insights and conclusions regarding the correlations. However, three or four (for the four correlation inequality) lists of measured or counterfactual data must be identified in a physical situation for (11) to be applicable. A method to accomplish this in the case of the usual experimental data is given below, and is the principle subject of this paper.

The situation when counterfactuals are used, as in Bell's gedankenexperiment, is more complicated [2]. If nonlocal interactions strong enough to change the data are assumed, a delayed choice experiment in which A is measured before B and B', still leads to three data lists, but consideration of setting A' as well as A before B and B', leads to six lists and inapplicability of the four correlation inequality. A method for circumventing this difficulty will be outlined in Sec. 4, and will be treated in more detail in a future paper.

## 1.3 Requirement on Experimental Data

In terms of the analysis above, which applies in a precisely parallel manner to the four correlation inequality, the common belief that experiments violate Bell's inequalities may now be understood. *While Bell's inequality appears to be a relation among statistics i.e. correlations, it is in fact an arithmetic identity that has been applied to statistics.* The correlations that have been substituted into Bell's inequalities have been computed among eight data lists [17,18] (for the four correlation inequality) rather than the four lists implied in its derivations. This is partly due to the fact that only one measurement per particle from a particle-pair may be obtained. Each correlation requires two data lists from a series of particle pairs at fixed detector settings. Independent trials at each pair of detector settings implies that the crucial factoring step used in (4) and (10) no longer holds. The inequality may now be violated by fluctuating correlation estimates, or by *assumed* ensemble average correlations that cannot result from actual cross-correlations of data lists for which the factoring step is automatically obeyed. That the consequences of this have not been understood previously, is apparently due to the failure to discover that Bell's inequality holds for more general processes than those spatially stationary ones used to derive it originally. *For the special case of a spatially stationary stochastic process, it is appropriate to determine the correlations at four angular differences in independent runs, and to insert their limiting values in Bell's inequality.* However, for a general non-stationary process, the factoring condition must be heeded so that the correlations depend on values of data occurring together. A procedure illustrating this, that has no effect on the correlations ordinarily measured, is specified in the next section.

## 2. A NEW EXPERIMENT

### 2.1 Procedure for Data Correlation

Correlations for three correlation Bell's inequalities will be treated first, followed by those for four correlations. The method does not conform to the specifications of Bell since data is collected in independent trials. In the final section of the paper, the probability implications of Bell's more stringent assumption will be briefly considered.

The procedure consists of measurement of $A(\theta_A)$ and $B(\theta_B)$, and then $A'(\theta_{A'})$ with $B(\theta_B)$ in two runs, so as to accumulate $N$ pairs of $\pm 1$'s, or two lists for each correlation. The data may then be made consistent with the factoring condition if data pairs of the second A'-B run are reordered so that the sequence of B-values is the same as in the first run while the sequence of A'-values is the same as occurred experimentally. This is possible since the probability for observation of $B = +1$ and $B = -1$ equals 1/2 in both the A-B and A'-B data sequences, and the sequence of B-values is random for each. The error from the mean fraction of +1's in N trials equals $.5/\sqrt{N}$, so that in the limit N, the fraction of $\pm 1$'s in the two lists of B-values is the same, allowing them to be matched. The value of correlation $\langle A'B \rangle$ is unaffected by changing the order of summation used in computing it. As shown below, the resulting correlation $\langle AA' \rangle$ may be predicted from the sum of correlations $\langle AA'|B=1 \rangle$ and $\langle AA'|B=-1 \rangle$ each weighted by 1/2, and using the probabilities given by quantum mechanics. The result may be checked against experimental data.

A similar procedure may be carried out in the case of four correlations. Three experiments must be run that generate six data lists for the variable pairs (A, B), (A', B), and (A, B'). In the limit of infinite list length, a sequence of pairs (A', B) may be obtained by reordering, so that the sequence of B-values is the same as that for (A, B). The procedure is repeated to obtain a sequence of pairs (A, B') such that the *A*-values match those in the previously existing A-B sequence. Once the appropriate variables have common values, the six data lists are effectively reduced to four, and the correlation of the pairs (A'B') may be computed.

### 2.2 Noncommutation Issues

It may be noted that the data matching procedures suggested bypass questions of noncommutativity among certain of the variables. In the notation used, primed variables do not commute with unprimed variables having the same letter name. Quantum mechanically this refers to simultaneous measurements on the same particle. But since in the prescribed procedure, noncommuting variables such as *A* and *A'* are measured with *B* in separate experiments, there is no conflict with quantum mechanical principles. *A* and *A'* are each correlated with *B*, and their correlation with each other is conditional on fixed *B* so that it can be non zero.

## 2.3 Theoretical Expressions for Three Correlations

The correlation of *A* and *B* resulting from the run of a correlation experiment may be written

$$\langle AB \rangle = \sum_{A,B} A(\theta_A) B(\theta_B) p(A(\theta_A), B(\theta_B)) \quad , \tag{12}$$

where $p(A(\theta_A), B(\theta_B))$ is a joint probability density given by quantum mechanics. The measurements of *A* and *B* involve different particles and therefore commute. Eq. (12) may also be written in terms of conditional probabilities:

$$\langle AB \rangle = \sum_{B,A} A(\theta_A) B(\theta_B) p(A(\theta_A)/B(\theta_B)) p(B(\theta_B)) \quad . \tag{13}$$

Similarly, a correlation using variables *A'-B* from a second independent run may be written

$$\langle A'B \rangle = \sum_{A'B} A'(\theta_{A'}) B(\theta_B) p(A'(\theta_{A'})/B(\theta_B)) p(B(\theta_B)) \quad . \tag{14}$$

The correlation of *A* and *A'* conditional on *B* as defined by the procedure of section 2.1 is

$$\langle AA' \rangle = \sum_{A,A',B} A(\theta_A) A'(\theta_{A'}) p(A(\theta_A), A'(\theta_{A'})/B) p(B) \quad . \tag{15}$$

Since *A* and *A'* are measured with *B* in separate independent experiments, they are statistically independent except for their conditional dependence on *B*. (These experiments could even be carried out in different laboratories and the data combined afterward.) This insures their (conditional) statistical independence whether or not hidden variables are assumed. Their conditional joint probability must therefore factor, leading to

$$\langle AA' \rangle = \sum_{A,A',B} A(\theta_A) A'(\theta_{A'}) p(A(\theta_A)/B(\theta_B)) p(A'(\theta_{A'})/B(\theta_B)) p(B(\theta_B)) \quad . \tag{16}$$

All the probabilities in (16) are now known from quantum mechanics.

## 2.4 Theoretical Expressions for Four Correlations

The correlations of *A-B*, *A'-B*, and *A-B'* are obtained from different, statistically independent experimental runs, and may be computed as in (13) above. The interesting case is the correlation $\langle A'B' \rangle$ which is conditional on values for *A* and *B*. In general one may write:

$$\langle A'B' \rangle = \sum_{A,B,A',B'} A'B' p(A',B'/A,B) p(A,B) \quad , \tag{17}$$

where the angle dependence of the variables has been suppressed for simplicity. Since *A'* and *B'* are measured in different experiments, they are statistically independent except for their conditional coupling to *B* and *A* respectively. Then

$$\langle A'B' \rangle = \sum_{A,B,A',B'} A'B' p(A'/A,B) p(B'/A,B) p(A,B) \quad . \tag{18}$$

Finally, since *A'* is measured in a separate experiment (instead of *A*) with *B*, and *B'* in a separate experiment (instead of *B*) with *A*, one has

$$\langle A'B' \rangle = \sum_{A,B,A'B'} A'B' \, p(A'/B) p(B'/A) \, p(A,B) \quad . \tag{19}$$

## 2.5 Correlations Based on Quantum Mechanical Probabilities

Equations (16) and (19) are now in a form appropriate for the insertion of the quantum mechanical probabilities. These are:

$$p(A = -1 \mid B = +1) = p(A = +1 \mid B = -1) = \cos^2 \frac{\theta_A - \theta_B}{2}$$

$$p(A = -1 \mid B = -1) = p(A = +1 \mid B = +1) = \sin^2 \frac{\theta_A - \theta_B}{2} \tag{20}$$

$$p(B = +1) = p(B = -1) = \frac{1}{2}$$

Expressions for probabilities $p(A' \mid B)$ and $p(B' \mid A)$, etc., are the same as (20) with appropriately relabeled angles.

### 2.5.1 Three Correlations

The result for $\langle AA' \rangle$ from (16) may be computed to be

$$\langle AA' \rangle = \cos(\theta_A - \theta_B)\cos(\theta_{A'} - \theta_B) \quad . \tag{21}$$

The correlation of $A'$ with $B$ is

$$\langle A'B \rangle = -\cos(\theta_{A'} - \theta_B) \quad , \tag{22}$$

and that of $A$ with $B$ is

$$\langle AB \rangle = -\cos(\theta_A - \theta_B) \quad . \tag{23}$$

### 2.5.2 Four Correlations

The interesting correlation is that of $A'$ with $B'$ which one may compute (after some algebra) to be:

$$\langle A'B' \rangle = -\cos(\theta_{A'} - \theta_B)\cos(\theta_{B'} - \theta_A)\cos(\theta_A - \theta_B) \quad . \tag{24}$$

The others are given by the negative cosine of angular differences as in the three correlation case. It should be noted that (21) is preceded by a plus sign while (24) is preceded by a minus sign. These differing signs are connected with the fact that $A$ and $A'$ lie on the same side of the apparatus, while $A'$ and $B'$ lie on opposite sides. Both correlations are defined in the experiment in spite of the fact that $A'$ and $B'$ commute while $A$ and $A'$ do not.

## 2.6 Satisfaction of Bell's Inequalities

Bell's inequalities for three and four correlations are:

$$|\langle AB \rangle - \langle A'B \rangle| + \langle AA' \rangle \leq 1 \tag{25}$$

$$|\langle AB \rangle + \langle AB' \rangle| + |\langle A'B \rangle - \langle A'B' \rangle| \leq 2 \tag{26}$$

Equations (21), (22) and (23) satisfy (25). Similarly, (24) satisfies (26) along with the usual correlations for the directly interacting observables.

For comparison, it is interesting to consider $\langle AA'\rangle$ in the case of three variables without invoking the matching condition. Now $\langle AA'\rangle = 0$ so that (25) is easily violated by the choice of 0 and $\pi$ for angles $\theta_A - \theta_B$ and $\theta_{A'} - \theta_B$, respectively. The correlation is not manifested when the matching condition is ignored, and the ensuing violation of Bell's inequality indicates that the resulting correlations are numerically unattainable by cross-correlating any three (infinite) data lists.

## 3. SIMULATIONS

Partial simulations have been carried out as a cross check on (21) and (24). The simulation for the three correlation case is based on (16) which may be rewritten as the average of conditional correlations of A and A' for B = +1 and B = -1:

$$\langle AA'\rangle = E\{A(\theta_A)A'(\theta_{A'})|B = +1\}\frac{1}{2} + E\{A(\theta_A)A'(\theta_{A'})|B = -1\}\frac{1}{2}\quad. \tag{27}$$

The expressions in (20) were used in conjunction with a random number generator to carry out the simulation using Mathematica. If when B = +1, A is set equal to -1 when the number generated is less than $\cos^2(\theta_A - \theta)/2$, and +1 otherwise; and when B = -1, A is set equal to -1 when the number generated is less than $\sin^2(\theta_A - \theta)/2$ and +1 otherwise, the conditional probabilities of (20) for A with respect to B will be realized. The expectations in (27) have been computed using this method based on $10^4$ trials at each pair of angle values. The result for $\theta_B = 0$ for values of $\theta_A$ and $\theta_{A'}$ from 0 to $\pi$ is shown in Fig. 2a. The correlation (21) is plotted over the same region in Fig. 2b.

The four correlation result may be simulated in a similar fashion.

## 4. BELL'S GEDANKENEXPERIMENT

A method for complete experimental realization of all correlations required for the applicability of Bell's inequality was described in the sections above. Such a realization appears to be as close to Bell's intent as can be achieved while satisfying the condition that all data are experimentally measured. However, Bell required that alternatives A' and B' be evaluated *for each experimental trial* with fixed values for measurements A, B, and $\lambda$, even though A and A' and B and B' are mutually exclusive alternatives and noncommuting observables.

The classical meaning of this is that multiple experiments must be performed with hidden variables fixed for each experimental trial, and with results at different detector settings measured. In the three variable case, this implies a joint density p(A, A', B) that does not allow the factoring step employed in going from (15) to (16) above. Similarly, in the four variable case the factoring of the joint density used in going from (17) to (19) is no longer justified. The resulting densities, marginal densities, and corresponding marginal correlations are in general different from those obtained above. Even though such correlations may be defined mathematically, there appears to be no way to measure them. They should not be assigned conventional quantum mechanical values, because the latter describe correlations for realizable experiments that should not be applied to *qualitatively different* unrealizable experiments. Further, if such correlations can be mathematically constructed, they cannot have the usually assigned quantum mechanical values because these violate Bell's inequality showing that they are inconsistent with any data that can possibly exist.

Cohen [19,20] has given a prescription for finding (positive) joint densities for multiple *continuous* random variables satisfying given conditions on the marginal statistics irrespective of whether the variables commute. In general, it is found that the marginal statistics do not uniquely determine joint densities.

The situation appears to be similar for the case of the discrete variables whose correlations are the subject of Bell's theorem. Quantum mechanics provides joint probability densities for spin measurements on the singlet state, and the conditional densities (20) immediately follow from probability theory. These densities depend on outcomes and relative angles, information considered to be nonlocal when shared in common between the two detectors. If a measurement outcome for B is given, the outcome for A($_A$) may be computed via a hidden variable simulation of the conditional probabilities from the nonlocal information. But if the outcome A($_A$) may be determined, then so may the alternate, equivalent outcome A'($_{A'}$), and therefore, the three correlations. The latter are nonstationary in angle coordinates and satisfy Bell's inequality. A similar solution may be found for the four variable inequality. Thus, even when nonlocality is assumed, quantum mechanical probabilities are consistent with three and four variables as appropriate, and the applicability of the associated Bell's inequalities follows. Of course, there is no known possible experimental method to confirm the correlations involving counterfactuals, although the measurable correlations are the same as before. The details of this solution and other analysis will be given in a future article.

## 5. CONCLUSION

It has been widely accepted that experimental data violate Bell's inequality in an unambiguous manner independently of theoretical assumptions. This ultimately implies that all correlations occurring in the inequality are measured experimentally. An experimental procedure has been described that allows all correlations to be measured in a way consistent with the factoring condition central to Bell's inequalities both in the three and four variable cases. The correlations are then computed from quantum mechanical probabilities. The resulting set of correlations exemplifies a process that is not wide sense stationary, and satisfies Bell's inequality. It should be observed that spatial stationarity is inconsistent with the noncommutation of quantum mechanical spin measurements.

Although wide sense stationarity was assumed in the historical derivation of Bell's theorem, it is not a necessary condition for the validity of Bell's inequality which holds for any process, random or deterministic, that produces appropriate lists of data. The violation of Bell's inequality by a correlation function of real data assumed to result from a stationary process, implies that the correlation cannot result from such a simple process. However, a more general nonstationary one is not necessarily ruled out.

**ACKNOWLEDGEMENT**

I am indebted to Michael J. Steiner for critical comments on the manuscript, and to Leon Cohen for informing me of the existence of reference [19]. Finally, I am indebted to Albert W. Saenz for stimulating conversations on subjects relating to this paper.

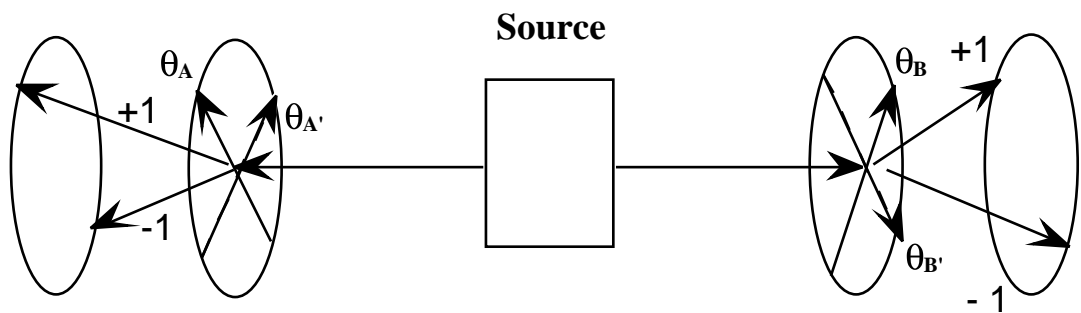

FIG. 1  Particles in singlet state are emitted in opposite directions and detected by Stern-Gerlach apparatus with orientations θ_A and θ_B, and alternate settings θ_A' and θ_B'.

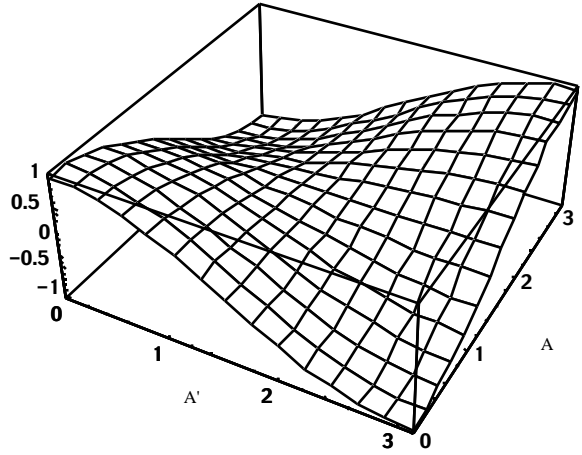

Fig 2a

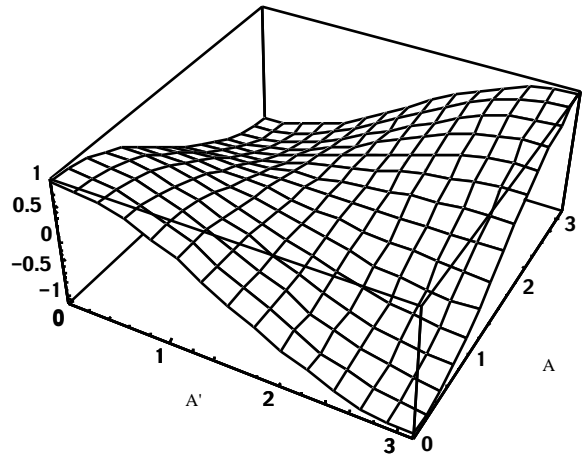

Fig. 2b

FIG. 2  (a) Simulation of Eq. (27) with $\phi_B = 0$. (b) Plot of Eq. (21) with $\phi_B = 0$.